# New Model of One-dimensional Completed Scattering and the Problem of Quantum Nonlocality

N. L. Chuprikov

*Department of Theoretical Physics, Tomsk State Pedagogical University, 634041, Tomsk, Russia*

**Abstract.** The origin of nonlocality in quantum mechanics (QM) is analyzed from the viewpoint of our new model of a one-dimensional (1D) completed scattering. Our study of quantum nonlocality complements those carried out by Volovich and Khrennikov. They pointed to an unphysical character of nonlocality in Bell's theorem whose context does not contain the very structure of the space-time. However, there is another reason leading to nonlocality in QM. The existing model of a 1D completed scattering evidences that QM, as it stands, even with a proper space-time context, contradicts special relativity. By our model this scattering process represents an entanglement of two coherently evolved alternative sub-processes, transmission and reflection; whose characteristics are measured well after the scattering event. Quantum nonlocality appears in this problem due to the inconsistency of the superposition principle with the corpuscular properties of a particle. It can take part only in one of the sub-processes. However the superposition principle allows introducing observables common for them. In the fresh wording, this principle must forbid introducing observables for entangled states.

## INTRODUCTION

Quantum nonlocality is one of the long-lived problems in the modern physics. By the well known Einstein-Podolsky-Rosen (EPR) [1] and Bell [2] theorems as well as by the experiment (see [3]), for compound quantum systems quantum mechanics predicts nonzero correlations between two space-like separated events.

From the practical point of view, this prediction of quantum mechanics is of promise, since it opens new perspectives in developing the various forms of information technology (see, e.g., [4]). However, from the theoretical viewpoint, the phenomenon of quantum nonlocality is an undesirable "guest" in physics; for it contradicts the principles of special relativity and, as a consequence, it violates the continuity in science.

An important step in resolving the problem of quantum nonlocality has been made by Igor Volovich and Andrei Khrennikov (see [5-9]) pointed to the fact that Bell's proof, which says about nonzero correlations between two space-like separated events, does not include the very structure of the space-time. That is, Bell's theorem has an improper basis to judge the (non)locality of quantum mechanics. As regards the existing experimental evidences in favor of quantum nonlocality, "…since detectors of particles are obviously located somewhere in space it shows that loop-holes are unavoidable in experiments aimed to establish a violation of Bell's inequalities... If the distance between the spatial regions where these detectors are located is large enough, then correlations between two measurements, performed simultaneously in these regions, are zero" [8].

However, the point is that the existing quantum mechanics leads to nonlocality even with a proper inclusion of the space-time structure. For example, the textbook quantum model of a 1D completed scattering, though equipped with the space-time structure, is non local (for details see [10,11]). This fact shows explicitly that quantum mechanics, as it stands, contradicts special relativity not only due to Bell's theorem.

## ABOUT NONLOCALITY IN THE EXISTING MODEL OF A ONE-DIMENSIONAL COMPLETED SCATTERING

As is known (see reviews [12-18]), during the last three decades a 1D completed scattering has been in the focus of the intensive debate on the so-called tunneling time problem, without reaching any consensus. Solving this problem has not been aimed at proving the existence of nonlocality in quantum mechanics. However, this phenomenon has arisen in all the existing approaches to the tunneling time problem.

The well-known group- and dwell-time concepts [12-18] are not exceptions. Like others they lead to unrealistic tunneling times. As was shown, for a transmitted particle the group and dwell times may be anomalously short or even negative by value. So that, the existing quantum-mechanical description of the scattering process, in fact, contradicts special relativity.

One can show that, in addition to nonlocality, the existing model of a 1D completed scattering is inconsistent, in principle (for more details see [10,11]). The fact that, in quantum mechanics, nonlocality and inconsistency accompany each other can be demonstrated in the case of the Bohmian variant of quantum mechanics.

Indeed (see [19]), the Bohmian model of the process predicts that the fate of any incident particle (to be transmitted or to be reflected by the barrier) must depend on the coordinate of its starting point. However, this property is evident to contradict the main principles of quantum mechanics, since any starting particle must have both the possibilities, irrespective of the location of its starting point.

Note that the position of the critical spatial point to separate the starting regions of to-be-transmitted and to-be-reflected particles depends on the shape of the potential barrier (though the barrier is located at a considerable distance from the particle's source). Thus, the "causal" trajectories of transmitted and reflected particles, introduced in the Bohmian mechanics, are, in fact, non-causal.

## A 1D COMPLETED SCATTERING AS AN ENTANGLEMENT OF TRANSMISSION AND REFLECTION

In a new model (see [10,11]) of a 1D completed scattering, for a particle impinging a symmetrical potential barrier from the left, we show that the (full) stationary wave function $\Psi_{full}(x;E)$ to describe this process can be uniquely presented in the form,

$\Psi_{full}(x;E) = \Psi_{tr}(x;E) + \Psi_{ref}(x;E)$, where $\Psi_{tr}(x;E)$ and $\Psi_{ref}(x;E)$ are solutions to the stationary Schrödinger equation; $E$ is the particle's energy. These functions allow us to retrace the time evolution of the (to-be-)transmitted and (to-be-)reflected subensembles of particles at all stages of scattering.

As was shown in [10,11], for any symmetric potential, $\Psi_{ref}(x;E)$ is an odd function with respect to the midpoint of the barrier region - $x_c$; i.e., $\Psi_{ref}(x_c;E) = 0$ for any value of $E$. This means that, in the case of reflection, particles impinging the barrier from the left do not enter the spatial region $x > x_c$.

Let $\psi_{ref}(x;E)$ denotes the wave packet to describe the subensemble of such particles: $\psi_{ref}(x;E) = \Psi_{ref}(x;E)$ for $x \leq x_c$ and $\psi_{ref}(x;E) \equiv 0$ for $x > x_c$. Then the subensemble of transmitted particles to impinge the barrier from the left and then passes through the barrier, without reflection and without violating the continuity equation at the midpoint $x_c$, will be described by the function $\psi_{tr}(x;E)$, where $\psi_{tr}(x;E) = \Psi_{full}(x;E) - \psi_{ref}(x;E)$; that is, $\psi_{tr}(x;E) = \Psi_{tr}(x;E)$ for $x \leq x_c$ and $\psi_{tr}(x;E) \equiv \Psi_{full}(x;E)$ for $x > x_c$. This property of $\psi_{tr}(x;E)$ results from the fact that the solutions $\Psi_{tr}(x;E)$ and $\Psi_{full}(x;E)$ have the same probability current density, and $\Psi_{tr}(x_c;E) = \Psi_{full}(x_c;E)$.

It is evident that wave packets $\psi_{tr}(x,t)$ and $\psi_{ref}(x,t)$ formed, respectively, from $\psi_{tr}(x;E)$ and $\psi_{ref}(x;E)$ are everywhere continuous, at any instant of time, and evolve with constant norms. It is important that the scalar product $\langle \psi_{tr}(x,t) | \psi_{ref}(x,t) \rangle$ is a purely imagine quantity to diminish when $t \to \infty$. By our approach, namely $\psi_{tr}(x,t)$ and $\psi_{ref}(x,t)$ (either includes only one incoming and only one outgoing wave packet) describe the time evolution of the (to-be-)transmitted and (to-be-)reflected subensembles of particles at all stages of scattering. On the macroscopic scales, particles of each subensemble have a common fate (or, history). Each subensemble moves along its own macroscopic (branchless) channel.

So, the superposition of two solutions of the Schrödinger equation, $\Psi_{tr}(x;E)$ and $\Psi_{ref}(x;E)$, is equivalent to that of $\psi_{tr}(x,t)$ and $\psi_{ref}(x,t)$ to describe transmission and reflection for a particle impinging the barrier from the left:
$$\Psi_{full}(x,t) = \Psi_{tr}(x,t) + \Psi_{ref}(x,t) = \psi_{tr}(x,t) + \psi_{ref}(x,t).$$
For any value of $t$, $\langle \psi_{tr}(x,t) | \psi_{tr}(x,t) \rangle = T = const$; $\langle \psi_{ref}(x,t) | \psi_{ref}(x,t) \rangle = R = const$, and $\langle \Psi_{full}(x,t) | \Psi_{full}(x,t) \rangle = T + R = 1$. Both the alternative sub-processes are macroscopically distinct at the final stage of scattering. So that $\Psi_{full}(x,t)$ should be considered as a Schrödinger's cat state: a 1D completed scattering to imply for a particle two macroscopically distinct alternative possibilities – to be transmitted or to be reflected by the barrier - represents an entanglement of two sub-processes during which these two possibilities are actualized.

The study of the temporal aspects of a 1D completed scattering, carried out on the basis of $\psi_{tr}(x,t)$ and $\psi_{ref}(x,t)$ (see [10,11]), shows that either subensemble behaves causally, without superluminal velocities. What is important is that, for each subensemble, the time spent on the average by particles in the barrier region can be measured by means of the well-known Larmor-clock procedure which is performed well after the scattering event, i.e., without demolishing the scattering process itself. In principle, this procedure to exploit the internal degree of freedom of a particle allows a non-demolishing scanning of either sub-process in any spatial interval.

As is shown in [10,11], the observed tunneling (Larmor) time is the well-known dwell time. As regards the group and other known concepts of tunneling time, they seem to have no physical sense for a scattering particle; for there is no (even *gedanken*) experiment to allow non-demolishing measurements of these quantities. Any particular point of wave packets cannot be used, as a representative of a particle, being in an entangled state, for timing its motion in the barrier region; it seems to be impossible to track up experimentally the motion of any chosen particular point of the (to-be-)transmitted (or (to-be-)reflected) wave packet at all stages of scattering.

## "MACROSCOPIC REALISM" AND THE SUPERPOSITION PRINCIPLE FOR PURE ENTANGLED STATES

Note, the above decomposition of the full (entangled) state of a particle is the only way to explain a 1D completed scattering. It is meaningless to introduce one-particle's observables for the whole ensemble of scattering particles. For all the measurements with the Larmor-clock procedure are performed well after the scattering event (when there is no interference between $\psi_{tr}(x,t)$ and $\psi_{ref}(x,t)$), i.e., for transmitted or reflected particles separately.

One has to take into account that a particle is an indivisible object, and it cannot simultaneously take part in two (or several) macroscopically distinct sub-processes. This means that the superposition principle must distinguish, on the conceptual level, between pure entangled states and pure unentangled (branchless) states. To reconcile quantum mechanics with classical one, it must forbid introducing observables for entangled states of micro-objects. Neglecting this rule leads to nonlocality.

We have to stress that a new model of a 1D one-particle's completed scattering contains the following three ingredients.

(1) *Macrorealism.* A scattering particle which has available to it two macroscopically distinct states is at any given time in a definite one of those states.

(2) *Non-invasive measurability.* By means of the Larmor-clock procedure it is possible to examine experimentally the average one-particle's characteristics for either sub-process, without any effect on the particle's state itself or on its subsequent dynamics.

(3) *Induction.* The properties of either subensemble are determined uniquely by initial conditions for the whole ensemble of particles.

Note that these ingredients coincide in fact with the requirements suggested by Leggett [20], which must be inherent to any 'macro-realistic' theory. Thus, our model of a 1D completed scattering is a 'macro-realistic' one. Of great importance is also to stress that, despite the interference between $\psi_{tr}(x,t)$ and $\psi_{ref}(x,t)$, the number of particles in either subensemble is constant during the whole scattering process. This interference does not influence the motion of particles in either channel; in fact, it is hidden when all measurements are performed well after the scattering event. At all stages of scattering a particle moves along a definite channel. There is no necessity to invoke 'environment' and 'decoherence', to reconcile micro- and macro-realism.

## CONCLUSION

So, by our approach, quantum nonlocality of entangled states is an artifact of the existing quantum theory. It appears in quantum mechanics due to the inconsistency of its superposition principle with the corpuscular properties of a particle. In the existing form it ignores the fact that a particle is indivisible object. To avoid the unphysical phenomenon of nonlocality of entangled states, quantum mechanics must forbid introducing observables for such states.

This receipt complements that obtained by Volovich and Khrennikov; to avoid the unphysical phenomenon of nonlocality for quantum processes in compound systems, one must not ignore the space-time structure in the mathematical model of the process.

And yet, entangled states do provide the basis for new perspectives in information technology. However, one has to bear in mind that a time-dependent entangled state does not describe a 'non-signaling' 'uncontrollable' process. Rather it describes the superposition of several coherent controllable sub-processes (signals), each evolving causally. To explain the time evolution of the interference pattern of this superposition, we must know the time evolution of every sub-process.